\documentclass[%
 reprint,
 amsmath,amssymb,
 aps,
prb,
]{revtex4-2}
\usepackage{braket}
\usepackage{graphicx}
\usepackage{dcolumn}
\usepackage{bm}
\usepackage{hyperref}

\begin{document}


\title{Using Crossed Andreev Reflection to Split Electrons}

\author{Austin Marga}
\author{Venkat Chandrasekhar}
\affiliation{Department of Physics and Astronomy, Northwestern University, 2145 Sheridan Road, Evanston, IL 60208, USA}
\email{v-chandrasekhar@northwestern.edu}

\date{\today}

\begin{abstract}
 Mesoscopic systems possess shot noise in their currents due to the quantization of the conducting quasiparticles. Measurements of this shot noise are useful to study phenomena that do not manifest themselves in standard conductance or resistance measurements, such as the statistics of the conducting quasiparticles or quantum entanglement via Bell tests \cite{blanter_shot_2000,zukowski_realizable_1997}. The corresponding particle statistics can be determined via two particle quantum interference experiments, such as the Hong-Ou-Mandel effect which demonstrates a bunching effect for bosons \cite{hong_measurement_1987} or an anti-bunching effect in fermions \cite{liu_quantum_1998,bocquillon_coherence_2013}. In superconducting proximity junctions, electrons incident on a superconductor can induce holes via crossed Andreev reflection (CAR) \cite{deutscher_coupling_2000} in spatially separated normal metal leads, where the resulting hole currents have nontrivial partition noise due to the four terminal configuration. These nonlocally generated currents, using a superconductor as a mesoscopic beam splitter, enable fabrication of mesoscopic analogs to quantum optics interferometers using metallic and superconducting films with multiport geometries.
\end{abstract}

\maketitle
Interference is an ubiquitous phenomenon in both classical and quantum physics.  In the quantum regime, certain effects are predicted that have no classical analog.  Well-known examples include single-photon interference, or the Hanbury Brown and Twiss effect \cite{brown_correlation_1956,grangier_experimental_1986}, and two-photon interference, or the Hong-Ou-Mandel effect \cite{hong_measurement_1987}, which demonstrates the bunching of photons that cannot be explained classically. Although both single-electron interference \cite{henny_fermionic_1999,oliver_hanbury_1999,bocquillon_electron_2012}, and two-electron interference \cite{liu_quantum_1998,bocquillon_coherence_2013} has been demonstrated, such experiments are not easy, primarily due to the difficulty of realizing the equivalent of a beam splitter for electrons.  Here we show that a superconductor of dimensions comparable to the superconducting coherence length $\xi$ can act as a beam splitter for quasiparticles in normal metal wires connected to it through the process of crossed Andreev reflection (CAR) \cite{deutscher_coupling_2000}, which coherently couples electrons in spatially separated normal metals through their mutual interaction with the superconductor.  This would enable observation of single-electron and two-electron interference in superconducting/normal-metal hybrid devices.

\begin{figure}
    \centering
    \includegraphics[width=0.4\linewidth]{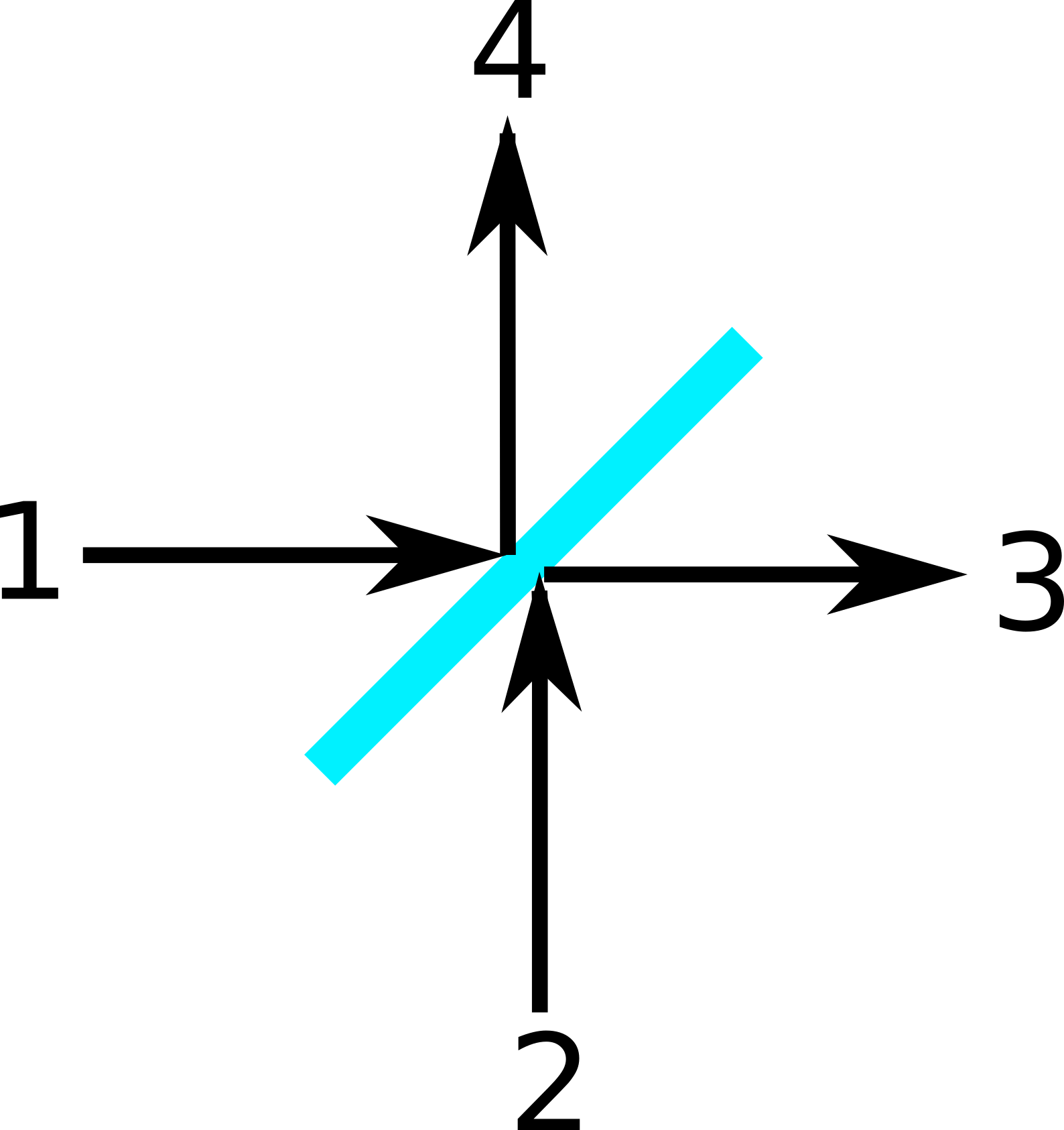}
    \caption{Schematic of the Hong-Ou-Mandel two-photon interference effect.}
    \label{fig:figure1}
\end{figure}
It is instructive to first review the Hong-Ou-Mandel effect for photons \cite{gerry_introductory_2004}.  Consider then a simple optical beam splitter shown in Fig. \ref{fig:figure1}. Photons incident from paths 1 and 2 on this beam splitter are each partially transmitted and partially reflected. The incoming photon along path 1 is transmitted to 3 with a coefficient $t$ and reflected into 4 with coefficient $r$. Similarly, the incoming photon in 2 is transmitted to 4 with coefficient $t$ and reflected into 3 with coefficient $r$. The coefficients $t$ and $r$ are in general complex quantities. This can be represented in a scattering matrix \cite{zeilinger_general_1981,prasad_quantum_1987}:
\begin{equation}
    \begin{pmatrix}
        \hat{a}_1^\dagger
        \\
        \hat{a}_2^\dagger
    \end{pmatrix} =\begin{pmatrix}
        t & r \\
        r & t
    \end{pmatrix}
    \begin{pmatrix}
        \hat{a}_3^\dagger
        \\
        \hat{a}_4^\dagger
    \end{pmatrix}
\end{equation}If we denote the initial state of the system by $\ket{i} = \hat{a}_1^\dagger\hat{a}_2^\dagger\ket{0} = \ket{1100}$ where $\ket{0} =\ket{0000}$ is the product of the empty state for each channel, and the final state as $\ket{f}$, then the state after the photons traverse the beam splitter can be represented in terms of the outgoing states of $\hat{a}_3^\dagger$ and $\hat{a}_4^\dagger$. 
\begin{equation}
    \ket{i} = \hat{a}_1^\dagger\hat{a}_2^\dagger\ket{0} \to (t\hat{a}_3^\dagger +r\hat{a}_4^\dagger)(r\hat{a}_3^\dagger +t\hat{a}_4^\dagger)\ket{0} = \ket{f}
\end{equation}
Requiring this transformation to be unitary due to number conservation of photons, we obtain the conditions $|t|^2 + |r|^2=1$ and $r^* t + t^*r=0$.  Setting $t\rightarrow te^{i\theta}$ and $r\rightarrow re^{i\phi}$, the second condition gives $\cos(\theta-\phi)=0$, whose simplest solution is $\theta-\phi=\pi/2$. Arbitrarily setting $\theta=0$, and considering the specific case of a 50:50 beam splitter, we finally obtain $t=1/\sqrt{2}$, $r=i/\sqrt{2}$.

Putting these values into our expression for the state after the photons pass through the beam splitter, we obtain
\begin{multline*}
 (t\hat{a}_3^\dagger +r\hat{a}_4^\dagger)(r\hat{a}_3^\dagger +t\hat{a}_4^\dagger)\ket{0} \\ 
 =\frac{1}{2}(i\hat{a}_3^\dagger + \hat{a}_4^\dagger)(\hat{a}_3^\dagger + i \hat{a}_4^\dagger)\ket{0}=\frac{1}{2}(i\hat{a}_3^\dagger \hat{a}_3^\dagger + i\hat{a}_4^\dagger \hat{a}_4^\dagger)\ket{0}\\
 =\frac{i}{2}(\ket{0020} + \ket{0002})
\end{multline*}
so that after transmission through the beam splitter, there is equal probability of either both photons being along path 3 or both photons being along path 4, but zero probability that one photon would be found in path 3 while the other is found in path 4.  This is the boson (photon) bunching effect discovered experimentally by Hong, Ou, and Mandel \cite{hong_measurement_1987}, which requires that the incoming photons be identical (i.e., they have the same polarization). \par
The analysis for a 50:50 beam splitter for electrons is similar, except that we have anticommuting fermionic creation ($\hat{c}^\dagger$) and annihilation ($\hat{c}$) operators instead of the commuting bosonic ones.  The equivalent expression for fermionic two particle interference is then
\begin{align*}
 \frac{1}{2}(i\hat{c}_3^\dagger + &\hat{c}_4^\dagger)(\hat{c}_3^\dagger + i \hat{c}_4^\dagger)\ket{0} \\ 
 &=\frac{1}{2}(i\hat{c}_3^\dagger \hat{c}_3^\dagger - \hat{c}_3^\dagger \hat{c}_4^\dagger + \hat{c}_4^\dagger \hat{c}_3^\dagger + i\hat{c}_4^\dagger \hat{c}_4^\dagger)\ket{0} \\
 &=\ket{0011}   
\end{align*}
so that the only possibility is that one electron is found along path 3 while the other is found along path 4, since both electrons cannot be found along the same path due to the Pauli exclusion principle.  This is the fermionic anti-bunching effect demonstrated with continuous electron sources \cite{liu_quantum_1998} and single electron sources \cite{bocquillon_coherence_2013}. Of course, as with the photon case, this only applies if the electrons are identical, which in our case implies that they have the same spin orientation. 
\par
There is a growing interest in electronic quantum optics due to the scalability inherent to solid state systems (a nice discussion of electronic quantum optics is presented in \cite{roussel_electron_2017}). While beam splitters are a standard component in a quantum optics toolkit, electronic beam splitters are more difficult to realize. Most implementations to date use quantum point contacts (QPCs) in high mobility two-dimensional electron gases (2DEGs) where the transmittance and reflectance of the beam splitter can be tuned by a gate voltage \cite{oliver_hanbury_1999}. Others have used chiral edge states from quantum Hall devices in addition to QPCs to minimize backscattering \cite{henny_fermionic_1999, liu_quantum_1998}. This has been further refined by having electrons emitted from a quantum dot, which provides a source of single electrons, which then interfere while in the edge states of a 2DEG at a QPC \cite{bocquillon_electron_2012,bocquillon_coherence_2013}. QPC beam splitters have also been used in broader applications, such as realizing an electronic equivalent to the Mach-Zehnder interferometer \cite{ji_electronic_2003} and, recently, interfering anyonic quasiparticles with fractional statistics \cite{bartolomei_fractional_2020, de_electronic_2025,taktak_two-particle_2022,kundu_anyonic_2023}. However, QPCs are more difficult to use with more terminals, and quantum Hall systems require large magnetic fields and high electron mobility. Here, we propose using a superconductor as an electronic beam splitter. This will allow us to circumvent the need for large magnetic fields, allow for an arbitrary amount of leads, and significantly simplify device fabrication.
\par Beam splitters based on superconductor/normal-metal hybrid structures have been discussed \cite{torres_positive_1999} and experimentally realized \cite{hofstetter_cooper_2009,recher_andreev_2001}, but such devices had at most 2 normal metals in contact with the superconductor or separated by quantum dots.
\begin{figure}
    \centering
    \includegraphics[width=0.4\linewidth]{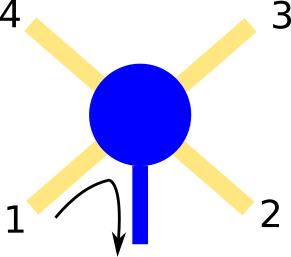}
    \caption{Schematic of a quasiparticle beam splitter based on crossed Andreev reflection.  Blue represents the superconductor while gold represents normal metal wires.  The diameter of the superconducting circle is of order of the superconducting coherence length $\xi$.  The superconducting line below the circle is to drain any current injected into the superconductor from leads 1 and 2, with the current from lead 1 portrayed above.}
    \label{fig:figure2}
\end{figure}
Here we consider a beam splitter based on a superconductor/normal-metal hybrid device with multiple, spatially separated normal metals. For simplicity, and to make a connection with the optical beam splitter, will will consider the four terminal configuration as shown in Fig. \ref{fig:figure2}, with a current applied to lead 1 and drained out the superconductor. An electron with a specific spin orientation incident on a normal-metal/superconductor (NS) interface in one of the normal metal leads can undergo a number of scattering processes at the NS interface:  it can be normally reflected, it can be Andreev reflected as a hole in the same lead, it can be Andreev reflected as a hole into one of the other leads (the process of CAR), or it can elastically co-tunnel into one of the other leads. We shall not discuss normal reflection here as it does not contribute to the cross correlations signals of interest.  We shall also not consider elastic co-tunneling (EC), as recent theory suggests when the inverse proximity effect is not strong, EC is much weaker than CAR \cite{tjernshaugen_crossed_2024}. Some experimental findings indicate that EC might even be entirely absent \cite{wei_positive_2010}. We will consider devices in this limit. The resulting CAR currents in spatially separated normal leads from lead 1 are nonlocal; they are not in the direction of conventional current flow.
\par
To simplify our discussion, we consider the case of zero temperature where the ground state in the normal leads is the vacuum state of the filled Fermi sea. A spin-up electron incident on the NS interface in lead 1 is then Andreev reflected as a spin-down hole into all the leads attached to the superconductor.  This process can be represented by 
\begin{equation}
    \hat{c}_{1\uparrow}^\dagger = \sum_{a=1}^{N=4}T_{1a}\hat{c}_{a\downarrow}
\end{equation}
Note that the analysis lends itself to an arbitrary amount of leads, but we will only consider the four terminal case. A similar scattering matrix as the photonic Hong-Ou-Mandel effect can be constructed as:
\begin{equation}
    \begin{pmatrix}
        \hat{c}_{1\uparrow}^\dagger \\
        \hat{c}_{2\uparrow}^\dagger \\
        \hat{c}_{3\uparrow}^\dagger \\
        \hat{c}_{4\uparrow}^\dagger \\
    \end{pmatrix} = \begin{pmatrix}
        T_{11} & T_{12} & T_{13} & T_{14} \\ 
        T_{21} & T_{22} & T_{23} & T_{24} \\ 
        T_{31} & T_{32} & T_{33} & T_{34} \\ 
        T_{41} & T_{42} & T_{43} & T_{44} \\ 
    \end{pmatrix}
    \begin{pmatrix}
        \hat{c}_{1\downarrow}\\
        \hat{c}_{2\downarrow}\\
        \hat{c}_{3\downarrow}\\
        \hat{c}_{4\downarrow}\\
    \end{pmatrix}
\end{equation}
where the diagonal elements $T_{aa}$ are the coefficients for local Andreev reflection, and the off-diagonal elements are the coefficients for CAR. The constraints on the scattering matrix $T$ due to unitarity are
\begin{equation}
    \sum_{b=1}^{4} T_{ab}T_{cb}^* = \sum_{b=1}^{4} T_{ba}^*T_{bc} = \delta_{ac}
\end{equation}
For example, the sum of the probabilities into all possible leads from lead 1 must be unity.
\begin{equation}
    |T_{11}|^2 + |T_{12}|^2 + |T_{13}|^2 + |T_{14}|^2 = 1
\end{equation}
Similarly, the sum of the probabilities into all possible leads from lead 2 must be unity.
\begin{equation}
    |T_{21}|^2 + |T_{22}|^2 + |T_{23}|^2 + |T_{24}|^2=1
\end{equation}
In general, the scattering matrix coefficients are complex quantities that depend on the transparency of the NS interfaces as well as the distance in the superconductor between NS interfaces. We expect the diagonal terms $|T_{aa}|^2$, which represent conventional Andreev reflection, to be larger than the off-diagonal CAR contributions.
\par
If we consider an electron incident only in lead 1, this configuration is then similar to that used to explore the Hanbury Brown and Twiss effect in the experiments on 2DEGs discussed earlier.  It is easy to confirm that the expectation value of holes in leads 3 and 4 after Andreev reflection is $\bra{f}\hat{c}_{3\downarrow} \hat{c}_{3\downarrow}^\dagger\ket{f} = |T_{13}|^2$ and $\bra{f}\hat{c}_{4\downarrow} \hat{c}_{4\downarrow}^\dagger\ket{f} = |T_{14}|^2$ respectively and that the cross correlation $\bra{f}\hat{c}_{3\downarrow} \hat{c}_{3\downarrow}^\dagger \hat{c}_{4\downarrow} \hat{c}_{4\downarrow}^\dagger\ket{f}$ vanishes.
\par As with the experiments on 2DEGs, however, implementing an experiment where only one electron impinges on the NS interface is difficult, as there are many quantized conductance channels present in mesoscopic leads \cite{blanter_shot_2000}, and experiments are usually performed by applying a voltage or sourcing a current.  In this case, the experimentally relevant measurement is the correlations in the noise in the current or voltage between different normal leads which can be encapsulated by the quantity
\begin{equation}
   g_{ab}= \frac{\langle \hat{c}_{a\downarrow}\hat{c}_{a\downarrow}^\dagger \hat{c}_{b\downarrow}\hat{c}_{b\downarrow}^\dagger\rangle - \langle \hat{c}_{a\downarrow}\hat{c}_{a\downarrow}^\dagger\rangle \langle \hat{c}_{b\downarrow}\hat{c}_{b\downarrow}^\dagger\rangle}{\langle \hat{c}_{a\downarrow}\hat{c}_{a\downarrow}^\dagger\rangle \langle \hat{c}_{b\downarrow}\hat{c}_{b\downarrow}^\dagger\rangle }
\end{equation}
where the angular brackets denote expectation values in the state $\ket{f}$ after Andreev reflection.  Using our previous results, we immediately obtain $g_{33} = |T_{13}|^{-2} -1$, $g_{44} = |T_{14}|^{-2} -1$,  and $g_{34}=-1$. $|T_{13}|^2$ and $|T_{14}|^2$ are both less than unity, so that the autocorrelations $g_{33}$ and $g_{44}$ are both positive while the cross correlation measure $g_{34}$ is perfectly anti-correlated.
\par Let us now consider the case of two-electron interference, one from lead $1$ and one from lead $2$. The initial state is then
\begin{equation}
    \ket{i} = \hat{c}_{1\uparrow}^\dagger \hat{c}_{2\uparrow}^\dagger \ket{0} = \ket{1100}
\end{equation}
After transforming our initial electron creation operators to hole creation operators when incident on the beam splitter, the average number of holes in lead $3$ is:
\begin{equation}
    \langle\hat{c}_{3\downarrow}\hat{c}_{3\downarrow}^\dagger \rangle = |T_{13}|^2 + |T_{23}|^2
\end{equation}
Similarly, the average number of holes in lead $4$ is:
\begin{equation}
    \langle \hat{c}_{4\downarrow}\hat{c}_{4\downarrow}^\dagger \rangle = |T_{14}|^2 + |T_{24}|^2
\end{equation}
If we next calculate the likelihood of a hole in both lead $3$ and lead $4$, we obtain:
\begin{widetext}
\begin{equation}
    \langle\hat{c}_{3\downarrow}\hat{c}_{3\downarrow}^\dagger \hat{c}_{4\downarrow}\hat{c}_{4\downarrow}^\dagger\rangle = |T_{14}|^2 |T_{23}|^2 + |T_{24}|^2 |T_{13}|^2 -( T_{14}T_{24}^*T_{13}^*T_{23}+T_{14}^*T_{24}T_{13}T_{23}^*)
\end{equation}
\end{widetext}
Our autocorrelation coefficients are 
\begin{equation}
    g_{33} = \Big(|T_{13}|^2 + |T_{23}|^2 \Big)^{-1}-1
\end{equation}
and 
\begin{equation}
    g_{44} = \Big(|T_{14}|^2 + |T_{24}|^2 \Big)^{-1}-1
\end{equation}
and the cross correlation coefficient is:
\begin{equation}
    g_{34} = -\frac{|T_{13}T_{14}^*-T_{23}T_{24}^*|^2}{\Big(|T_{13}|^2 +|T_{23}|^2\Big)\Big(|T_{14}|^2+|T_{24}|^2\Big)}\leq 0
\end{equation}
Since $g_{34}$ is a negative quantity in all realistic cases, there is a distinct, but not perfect, anti-correlative feature in measurement. This agrees with expectations. If one electron is incident from lead 1 and one from lead 2, there will be holes ejected with some weighted probabilities into any pair of leads. This is distinct from the strict anti-correlation found in the conventional electronic cases.
\par
In summary, beam splitters across sub-disciplines in physics are useful tools to investigate and utilize the quantum properties of incoming streams of particles \cite{zeilinger_general_1981}. The effects of particle statistics, either bosonic \cite{hong_measurement_1987}, fermionic \cite{liu_quantum_1998,bocquillon_coherence_2013}, or anyonic \cite{bartolomei_fractional_2020}, are highlighted in two-particle interference tests. Their mesoscopic equivalents, along with associated shot noise measurements, provide information in non-equilibrium systems not readily accessible by resistance or conductance measurements. Our proposed device possesses key advantages and points of interest over other electronic beam splitters. There is no need for a large magnetic field. The device fabrication is also simpler, as metallic films can be used compared to high-mobility GaAs crystalline films, and more elaborate devices with an arbitrary amount of ports can be constructed. The nonlocal CAR signals will be small, but the negative cross correlation coefficent provides a clear experimental signal for future work. Further extensions of the Andreev beam splitter can be used as a lumped element for further construction of mesoscopic interferometric devices. Potential applications could include various solid-state interferometers for tests of entanglement and quantum information \cite{campos_three-photon_2000,schaeff_experimental_2015,zukowski_realizable_1997,roussel_electron_2017}.
\begin{acknowledgments}
We would like to thank Sarvesh Upadhyay for the many fruitful discussions. This research was conducted with support from the National Science Foundation under Grant No. DMR-2303536.
\end{acknowledgments}

\bibliography{bib.bib}

\begin{thebibliography}{27}%
\makeatletter
\providecommand \@ifxundefined [1]{%
 \@ifx{#1\undefined}
}%
\providecommand \@ifnum [1]{%
 \ifnum #1\expandafter \@firstoftwo
 \else \expandafter \@secondoftwo
 \fi
}%
\providecommand \@ifx [1]{%
 \ifx #1\expandafter \@firstoftwo
 \else \expandafter \@secondoftwo
 \fi
}%
\providecommand \natexlab [1]{#1}%
\providecommand \enquote  [1]{``#1''}%
\providecommand \bibnamefont  [1]{#1}%
\providecommand \bibfnamefont [1]{#1}%
\providecommand \citenamefont [1]{#1}%
\providecommand \href@noop [0]{\@secondoftwo}%
\providecommand \href [0]{\begingroup \@sanitize@url \@href}%
\providecommand \@href[1]{\@@startlink{#1}\@@href}%
\providecommand \@@href[1]{\endgroup#1\@@endlink}%
\providecommand \@sanitize@url [0]{\catcode `\\12\catcode `\$12\catcode `\&12\catcode `\#12\catcode `\^12\catcode `\_12\catcode `\%12\relax}%
\providecommand \@@startlink[1]{}%
\providecommand \@@endlink[0]{}%
\providecommand \url  [0]{\begingroup\@sanitize@url \@url }%
\providecommand \@url [1]{\endgroup\@href {#1}{\urlprefix }}%
\providecommand \urlprefix  [0]{URL }%
\providecommand \Eprint [0]{\href }%
\providecommand \doibase [0]{https://doi.org/}%
\providecommand \selectlanguage [0]{\@gobble}%
\providecommand \bibinfo  [0]{\@secondoftwo}%
\providecommand \bibfield  [0]{\@secondoftwo}%
\providecommand \translation [1]{[#1]}%
\providecommand \BibitemOpen [0]{}%
\providecommand \bibitemStop [0]{}%
\providecommand \bibitemNoStop [0]{.\EOS\space}%
\providecommand \EOS [0]{\spacefactor3000\relax}%
\providecommand \BibitemShut  [1]{\csname bibitem#1\endcsname}%
\let\auto@bib@innerbib\@empty
\bibitem [{\citenamefont {Blanter}\ and\ \citenamefont {Büttiker}(2000)}]{blanter_shot_2000}%
  \BibitemOpen
  \bibfield  {author} {\bibinfo {author} {\bibfnamefont {Y.~M.}\ \bibnamefont {Blanter}}\ and\ \bibinfo {author} {\bibfnamefont {M.}~\bibnamefont {Büttiker}},\ }\href {https://doi.org/10.1016/S0370-1573(99)00123-4} {\bibfield  {journal} {\bibinfo  {journal} {Physics Reports}\ }\textbf {\bibinfo {volume} {336}},\ \bibinfo {pages} {1} (\bibinfo {year} {2000})}\BibitemShut {NoStop}%
\bibitem [{\citenamefont {Żukowski}\ \emph {et~al.}(1997)\citenamefont {Żukowski}, \citenamefont {Zeilinger},\ and\ \citenamefont {Horne}}]{zukowski_realizable_1997}%
  \BibitemOpen
  \bibfield  {author} {\bibinfo {author} {\bibfnamefont {M.}~\bibnamefont {Żukowski}}, \bibinfo {author} {\bibfnamefont {A.}~\bibnamefont {Zeilinger}},\ and\ \bibinfo {author} {\bibfnamefont {M.~A.}\ \bibnamefont {Horne}},\ }\href {https://doi.org/10.1103/PhysRevA.55.2564} {\bibfield  {journal} {\bibinfo  {journal} {Physical Review A}\ }\textbf {\bibinfo {volume} {55}},\ \bibinfo {pages} {2564} (\bibinfo {year} {1997})}\BibitemShut {NoStop}%
\bibitem [{\citenamefont {Hong}\ \emph {et~al.}(1987)\citenamefont {Hong}, \citenamefont {Ou},\ and\ \citenamefont {Mandel}}]{hong_measurement_1987}%
  \BibitemOpen
  \bibfield  {author} {\bibinfo {author} {\bibfnamefont {C.~K.}\ \bibnamefont {Hong}}, \bibinfo {author} {\bibfnamefont {Z.~Y.}\ \bibnamefont {Ou}},\ and\ \bibinfo {author} {\bibfnamefont {L.}~\bibnamefont {Mandel}},\ }\href {https://doi.org/10.1103/PhysRevLett.59.2044} {\bibfield  {journal} {\bibinfo  {journal} {Physical Review Letters}\ }\textbf {\bibinfo {volume} {59}},\ \bibinfo {pages} {2044} (\bibinfo {year} {1987})}\BibitemShut {NoStop}%
\bibitem [{\citenamefont {Liu}\ \emph {et~al.}(1998)\citenamefont {Liu}, \citenamefont {Odom}, \citenamefont {Yamamoto},\ and\ \citenamefont {Tarucha}}]{liu_quantum_1998}%
  \BibitemOpen
  \bibfield  {author} {\bibinfo {author} {\bibfnamefont {R.~C.}\ \bibnamefont {Liu}}, \bibinfo {author} {\bibfnamefont {B.}~\bibnamefont {Odom}}, \bibinfo {author} {\bibfnamefont {Y.}~\bibnamefont {Yamamoto}},\ and\ \bibinfo {author} {\bibfnamefont {S.}~\bibnamefont {Tarucha}},\ }\href {https://doi.org/10.1038/34611} {\bibfield  {journal} {\bibinfo  {journal} {Nature}\ }\textbf {\bibinfo {volume} {391}},\ \bibinfo {pages} {263} (\bibinfo {year} {1998})}\BibitemShut {NoStop}%
\bibitem [{\citenamefont {Bocquillon}\ \emph {et~al.}(2013)\citenamefont {Bocquillon}, \citenamefont {Freulon}, \citenamefont {Berroir}, \citenamefont {Degiovanni}, \citenamefont {Plaçais}, \citenamefont {Cavanna}, \citenamefont {Jin},\ and\ \citenamefont {Fève}}]{bocquillon_coherence_2013}%
  \BibitemOpen
  \bibfield  {author} {\bibinfo {author} {\bibfnamefont {E.}~\bibnamefont {Bocquillon}}, \bibinfo {author} {\bibfnamefont {V.}~\bibnamefont {Freulon}}, \bibinfo {author} {\bibfnamefont {J.-M.}\ \bibnamefont {Berroir}}, \bibinfo {author} {\bibfnamefont {P.}~\bibnamefont {Degiovanni}}, \bibinfo {author} {\bibfnamefont {B.}~\bibnamefont {Plaçais}}, \bibinfo {author} {\bibfnamefont {A.}~\bibnamefont {Cavanna}}, \bibinfo {author} {\bibfnamefont {Y.}~\bibnamefont {Jin}},\ and\ \bibinfo {author} {\bibfnamefont {G.}~\bibnamefont {Fève}},\ }\href {https://doi.org/10.1126/science.1232572} {\bibfield  {journal} {\bibinfo  {journal} {Science}\ }\textbf {\bibinfo {volume} {339}},\ \bibinfo {pages} {1054} (\bibinfo {year} {2013})}\BibitemShut {NoStop}%
\bibitem [{\citenamefont {Deutscher}\ and\ \citenamefont {Feinberg}(2000)}]{deutscher_coupling_2000}%
  \BibitemOpen
  \bibfield  {author} {\bibinfo {author} {\bibfnamefont {G.}~\bibnamefont {Deutscher}}\ and\ \bibinfo {author} {\bibfnamefont {D.}~\bibnamefont {Feinberg}},\ }\href {https://doi.org/10.1063/1.125796} {\bibfield  {journal} {\bibinfo  {journal} {Applied Physics Letters}\ }\textbf {\bibinfo {volume} {76}},\ \bibinfo {pages} {487} (\bibinfo {year} {2000})}\BibitemShut {NoStop}%
\bibitem [{\citenamefont {Brown}\ and\ \citenamefont {Twiss}(1956)}]{brown_correlation_1956}%
  \BibitemOpen
  \bibfield  {author} {\bibinfo {author} {\bibfnamefont {R.~H.}\ \bibnamefont {Brown}}\ and\ \bibinfo {author} {\bibfnamefont {R.~Q.}\ \bibnamefont {Twiss}},\ }\href {https://doi.org/10.1038/177027a0} {\bibfield  {journal} {\bibinfo  {journal} {Nature}\ }\textbf {\bibinfo {volume} {177}},\ \bibinfo {pages} {27} (\bibinfo {year} {1956})}\BibitemShut {NoStop}%
\bibitem [{\citenamefont {Grangier}\ \emph {et~al.}(1986)\citenamefont {Grangier}, \citenamefont {Roger},\ and\ \citenamefont {Aspect}}]{grangier_experimental_1986}%
  \BibitemOpen
  \bibfield  {author} {\bibinfo {author} {\bibfnamefont {P.}~\bibnamefont {Grangier}}, \bibinfo {author} {\bibfnamefont {G.}~\bibnamefont {Roger}},\ and\ \bibinfo {author} {\bibfnamefont {A.}~\bibnamefont {Aspect}},\ }\href {https://doi.org/10.1209/0295-5075/1/4/004} {\bibfield  {journal} {\bibinfo  {journal} {Europhysics Letters}\ }\textbf {\bibinfo {volume} {1}},\ \bibinfo {pages} {173} (\bibinfo {year} {1986})}\BibitemShut {NoStop}%
\bibitem [{\citenamefont {Henny}\ \emph {et~al.}(1999)\citenamefont {Henny}, \citenamefont {Oberholzer}, \citenamefont {Strunk}, \citenamefont {Heinzel}, \citenamefont {Ensslin}, \citenamefont {Holland},\ and\ \citenamefont {Schonenberger}}]{henny_fermionic_1999}%
  \BibitemOpen
  \bibfield  {author} {\bibinfo {author} {\bibfnamefont {M.}~\bibnamefont {Henny}}, \bibinfo {author} {\bibfnamefont {S.}~\bibnamefont {Oberholzer}}, \bibinfo {author} {\bibfnamefont {C.}~\bibnamefont {Strunk}}, \bibinfo {author} {\bibfnamefont {T.}~\bibnamefont {Heinzel}}, \bibinfo {author} {\bibfnamefont {K.}~\bibnamefont {Ensslin}}, \bibinfo {author} {\bibfnamefont {M.}~\bibnamefont {Holland}},\ and\ \bibinfo {author} {\bibfnamefont {C.}~\bibnamefont {Schonenberger}},\ }\href {https://doi.org/10.1126/science.284.5412.296} {\bibfield  {journal} {\bibinfo  {journal} {Science}\ }\textbf {\bibinfo {volume} {284}},\ \bibinfo {pages} {296} (\bibinfo {year} {1999})}\BibitemShut {NoStop}%
\bibitem [{\citenamefont {Oliver}\ \emph {et~al.}(1999)\citenamefont {Oliver}, \citenamefont {Kim}, \citenamefont {Liu},\ and\ \citenamefont {Yamamoto}}]{oliver_hanbury_1999}%
  \BibitemOpen
  \bibfield  {author} {\bibinfo {author} {\bibfnamefont {W.~D.}\ \bibnamefont {Oliver}}, \bibinfo {author} {\bibfnamefont {J.}~\bibnamefont {Kim}}, \bibinfo {author} {\bibfnamefont {R.~C.}\ \bibnamefont {Liu}},\ and\ \bibinfo {author} {\bibfnamefont {Y.}~\bibnamefont {Yamamoto}},\ }\href {https://doi.org/10.1126/science.284.5412.299} {\bibfield  {journal} {\bibinfo  {journal} {Science}\ }\textbf {\bibinfo {volume} {284}},\ \bibinfo {pages} {299} (\bibinfo {year} {1999})}\BibitemShut {NoStop}%
\bibitem [{\citenamefont {Bocquillon}\ \emph {et~al.}(2012)\citenamefont {Bocquillon}, \citenamefont {Parmentier}, \citenamefont {Grenier}, \citenamefont {Berroir}, \citenamefont {Degiovanni}, \citenamefont {Glattli}, \citenamefont {Plaçais}, \citenamefont {Cavanna}, \citenamefont {Jin},\ and\ \citenamefont {Fève}}]{bocquillon_electron_2012}%
  \BibitemOpen
  \bibfield  {author} {\bibinfo {author} {\bibfnamefont {E.}~\bibnamefont {Bocquillon}}, \bibinfo {author} {\bibfnamefont {F.~D.}\ \bibnamefont {Parmentier}}, \bibinfo {author} {\bibfnamefont {C.}~\bibnamefont {Grenier}}, \bibinfo {author} {\bibfnamefont {J.-M.}\ \bibnamefont {Berroir}}, \bibinfo {author} {\bibfnamefont {P.}~\bibnamefont {Degiovanni}}, \bibinfo {author} {\bibfnamefont {D.~C.}\ \bibnamefont {Glattli}}, \bibinfo {author} {\bibfnamefont {B.}~\bibnamefont {Plaçais}}, \bibinfo {author} {\bibfnamefont {A.}~\bibnamefont {Cavanna}}, \bibinfo {author} {\bibfnamefont {Y.}~\bibnamefont {Jin}},\ and\ \bibinfo {author} {\bibfnamefont {G.}~\bibnamefont {Fève}},\ }\href {https://doi.org/10.1103/PhysRevLett.108.196803} {\bibfield  {journal} {\bibinfo  {journal} {Physical Review Letters}\ }\textbf {\bibinfo {volume} {108}},\ \bibinfo {pages} {196803} (\bibinfo {year} {2012})}\BibitemShut {NoStop}%
\bibitem [{\citenamefont {Gerry}\ and\ \citenamefont {Knight}(2004)}]{gerry_introductory_2004}%
  \BibitemOpen
  \bibfield  {author} {\bibinfo {author} {\bibfnamefont {C.}~\bibnamefont {Gerry}}\ and\ \bibinfo {author} {\bibfnamefont {P.}~\bibnamefont {Knight}},\ }\href@noop {} {\emph {\bibinfo {title} {Introductory {Quantum} {Optics}}}}\ (\bibinfo  {publisher} {Cambridge University Press},\ \bibinfo {year} {2004})\BibitemShut {NoStop}%
\bibitem [{\citenamefont {Zeilinger}(1981)}]{zeilinger_general_1981}%
  \BibitemOpen
  \bibfield  {author} {\bibinfo {author} {\bibfnamefont {A.}~\bibnamefont {Zeilinger}},\ }\href {https://doi.org/10.1119/1.12387} {\bibfield  {journal} {\bibinfo  {journal} {American Journal of Physics}\ }\textbf {\bibinfo {volume} {49}},\ \bibinfo {pages} {882} (\bibinfo {year} {1981})}\BibitemShut {NoStop}%
\bibitem [{\citenamefont {Prasad}\ \emph {et~al.}(1987)\citenamefont {Prasad}, \citenamefont {Scully},\ and\ \citenamefont {Martienssen}}]{prasad_quantum_1987}%
  \BibitemOpen
  \bibfield  {author} {\bibinfo {author} {\bibfnamefont {S.}~\bibnamefont {Prasad}}, \bibinfo {author} {\bibfnamefont {M.~O.}\ \bibnamefont {Scully}},\ and\ \bibinfo {author} {\bibfnamefont {W.}~\bibnamefont {Martienssen}},\ }\href {https://doi.org/10.1016/0030-4018(87)90015-0} {\bibfield  {journal} {\bibinfo  {journal} {Optics Communications}\ }\textbf {\bibinfo {volume} {62}},\ \bibinfo {pages} {139} (\bibinfo {year} {1987})}\BibitemShut {NoStop}%
\bibitem [{\citenamefont {Roussel}\ \emph {et~al.}(2017)\citenamefont {Roussel}, \citenamefont {Cabart}, \citenamefont {Fève}, \citenamefont {Thibierge},\ and\ \citenamefont {Degiovanni}}]{roussel_electron_2017}%
  \BibitemOpen
  \bibfield  {author} {\bibinfo {author} {\bibfnamefont {B.}~\bibnamefont {Roussel}}, \bibinfo {author} {\bibfnamefont {C.}~\bibnamefont {Cabart}}, \bibinfo {author} {\bibfnamefont {G.}~\bibnamefont {Fève}}, \bibinfo {author} {\bibfnamefont {E.}~\bibnamefont {Thibierge}},\ and\ \bibinfo {author} {\bibfnamefont {P.}~\bibnamefont {Degiovanni}},\ }\href {https://doi.org/10.1002/pssb.201600621} {\bibfield  {journal} {\bibinfo  {journal} {physica status solidi (b)}\ }\textbf {\bibinfo {volume} {254}},\ \bibinfo {pages} {1600621} (\bibinfo {year} {2017})}\BibitemShut {NoStop}%
\bibitem [{\citenamefont {Ji}\ \emph {et~al.}(2003)\citenamefont {Ji}, \citenamefont {Chung}, \citenamefont {Sprinzak}, \citenamefont {Heiblum}, \citenamefont {Mahalu},\ and\ \citenamefont {Shtrikman}}]{ji_electronic_2003}%
  \BibitemOpen
  \bibfield  {author} {\bibinfo {author} {\bibfnamefont {Y.}~\bibnamefont {Ji}}, \bibinfo {author} {\bibfnamefont {Y.}~\bibnamefont {Chung}}, \bibinfo {author} {\bibfnamefont {D.}~\bibnamefont {Sprinzak}}, \bibinfo {author} {\bibfnamefont {M.}~\bibnamefont {Heiblum}}, \bibinfo {author} {\bibfnamefont {D.}~\bibnamefont {Mahalu}},\ and\ \bibinfo {author} {\bibfnamefont {H.}~\bibnamefont {Shtrikman}},\ }\href {https://doi.org/10.1038/nature01503} {\bibfield  {journal} {\bibinfo  {journal} {Nature}\ }\textbf {\bibinfo {volume} {422}},\ \bibinfo {pages} {415} (\bibinfo {year} {2003})}\BibitemShut {NoStop}%
\bibitem [{\citenamefont {Bartolomei}\ \emph {et~al.}(2020)\citenamefont {Bartolomei}, \citenamefont {Kumar}, \citenamefont {Bisognin}, \citenamefont {Marguerite}, \citenamefont {Berroir}, \citenamefont {Bocquillon}, \citenamefont {Plaçais}, \citenamefont {Cavanna}, \citenamefont {Dong}, \citenamefont {Gennser}, \citenamefont {Jin},\ and\ \citenamefont {Fève}}]{bartolomei_fractional_2020}%
  \BibitemOpen
  \bibfield  {author} {\bibinfo {author} {\bibfnamefont {H.}~\bibnamefont {Bartolomei}}, \bibinfo {author} {\bibfnamefont {M.}~\bibnamefont {Kumar}}, \bibinfo {author} {\bibfnamefont {R.}~\bibnamefont {Bisognin}}, \bibinfo {author} {\bibfnamefont {A.}~\bibnamefont {Marguerite}}, \bibinfo {author} {\bibfnamefont {J.-M.}\ \bibnamefont {Berroir}}, \bibinfo {author} {\bibfnamefont {E.}~\bibnamefont {Bocquillon}}, \bibinfo {author} {\bibfnamefont {B.}~\bibnamefont {Plaçais}}, \bibinfo {author} {\bibfnamefont {A.}~\bibnamefont {Cavanna}}, \bibinfo {author} {\bibfnamefont {Q.}~\bibnamefont {Dong}}, \bibinfo {author} {\bibfnamefont {U.}~\bibnamefont {Gennser}}, \bibinfo {author} {\bibfnamefont {Y.}~\bibnamefont {Jin}},\ and\ \bibinfo {author} {\bibfnamefont {G.}~\bibnamefont {Fève}},\ }\href {https://doi.org/10.1126/science.aaz5601} {\bibfield  {journal} {\bibinfo  {journal} {Science}\ }\textbf {\bibinfo {volume} {368}},\ \bibinfo {pages} {173} (\bibinfo {year} {2020})}\BibitemShut {NoStop}%
\bibitem [{\citenamefont {De}\ \emph {et~al.}(2025)\citenamefont {De}, \citenamefont {Boudet}, \citenamefont {Nath}, \citenamefont {Kapfer}, \citenamefont {Farrer}, \citenamefont {Ritchie}, \citenamefont {Roulleau},\ and\ \citenamefont {Glattli}}]{de_electronic_2025}%
  \BibitemOpen
  \bibfield  {author} {\bibinfo {author} {\bibfnamefont {A.}~\bibnamefont {De}}, \bibinfo {author} {\bibfnamefont {C.}~\bibnamefont {Boudet}}, \bibinfo {author} {\bibfnamefont {J.}~\bibnamefont {Nath}}, \bibinfo {author} {\bibfnamefont {M.}~\bibnamefont {Kapfer}}, \bibinfo {author} {\bibfnamefont {I.}~\bibnamefont {Farrer}}, \bibinfo {author} {\bibfnamefont {D.~A.}\ \bibnamefont {Ritchie}}, \bibinfo {author} {\bibfnamefont {P.}~\bibnamefont {Roulleau}},\ and\ \bibinfo {author} {\bibfnamefont {D.~C.}\ \bibnamefont {Glattli}},\ }\href {https://doi.org/10.1038/s41467-025-63308-2} {\bibfield  {journal} {\bibinfo  {journal} {Nature Communications}\ }\textbf {\bibinfo {volume} {16}},\ \bibinfo {pages} {8466} (\bibinfo {year} {2025})}\BibitemShut {NoStop}%
\bibitem [{\citenamefont {Taktak}\ \emph {et~al.}(2022)\citenamefont {Taktak}, \citenamefont {Kapfer}, \citenamefont {Nath}, \citenamefont {Roulleau}, \citenamefont {Acciai}, \citenamefont {Splettstoesser}, \citenamefont {Farrer}, \citenamefont {Ritchie},\ and\ \citenamefont {Glattli}}]{taktak_two-particle_2022}%
  \BibitemOpen
  \bibfield  {author} {\bibinfo {author} {\bibfnamefont {I.}~\bibnamefont {Taktak}}, \bibinfo {author} {\bibfnamefont {M.}~\bibnamefont {Kapfer}}, \bibinfo {author} {\bibfnamefont {J.}~\bibnamefont {Nath}}, \bibinfo {author} {\bibfnamefont {P.}~\bibnamefont {Roulleau}}, \bibinfo {author} {\bibfnamefont {M.}~\bibnamefont {Acciai}}, \bibinfo {author} {\bibfnamefont {J.}~\bibnamefont {Splettstoesser}}, \bibinfo {author} {\bibfnamefont {I.}~\bibnamefont {Farrer}}, \bibinfo {author} {\bibfnamefont {D.~A.}\ \bibnamefont {Ritchie}},\ and\ \bibinfo {author} {\bibfnamefont {D.~C.}\ \bibnamefont {Glattli}},\ }\href {https://doi.org/10.1038/s41467-022-33603-3} {\bibfield  {journal} {\bibinfo  {journal} {Nature Communications}\ }\textbf {\bibinfo {volume} {13}},\ \bibinfo {pages} {5863} (\bibinfo {year} {2022})}\BibitemShut {NoStop}%
\bibitem [{\citenamefont {Kundu}\ \emph {et~al.}(2023)\citenamefont {Kundu}, \citenamefont {Biswas}, \citenamefont {Ofek}, \citenamefont {Umansky},\ and\ \citenamefont {Heiblum}}]{kundu_anyonic_2023}%
  \BibitemOpen
  \bibfield  {author} {\bibinfo {author} {\bibfnamefont {H.~K.}\ \bibnamefont {Kundu}}, \bibinfo {author} {\bibfnamefont {S.}~\bibnamefont {Biswas}}, \bibinfo {author} {\bibfnamefont {N.}~\bibnamefont {Ofek}}, \bibinfo {author} {\bibfnamefont {V.}~\bibnamefont {Umansky}},\ and\ \bibinfo {author} {\bibfnamefont {M.}~\bibnamefont {Heiblum}},\ }\href {https://doi.org/10.1038/s41567-022-01899-z} {\bibfield  {journal} {\bibinfo  {journal} {Nature Physics}\ }\textbf {\bibinfo {volume} {19}},\ \bibinfo {pages} {515} (\bibinfo {year} {2023})}\BibitemShut {NoStop}%
\bibitem [{\citenamefont {Torrès}\ and\ \citenamefont {Martin}(1999)}]{torres_positive_1999}%
  \BibitemOpen
  \bibfield  {author} {\bibinfo {author} {\bibfnamefont {J.}~\bibnamefont {Torrès}}\ and\ \bibinfo {author} {\bibfnamefont {T.}~\bibnamefont {Martin}},\ }\href {https://doi.org/10.1007/s100510051010} {\bibfield  {journal} {\bibinfo  {journal} {The European Physical Journal B - Condensed Matter and Complex Systems}\ }\textbf {\bibinfo {volume} {12}},\ \bibinfo {pages} {319} (\bibinfo {year} {1999})}\BibitemShut {NoStop}%
\bibitem [{\citenamefont {Hofstetter}\ \emph {et~al.}(2009)\citenamefont {Hofstetter}, \citenamefont {Csonka}, \citenamefont {Nygård},\ and\ \citenamefont {Schönenberger}}]{hofstetter_cooper_2009}%
  \BibitemOpen
  \bibfield  {author} {\bibinfo {author} {\bibfnamefont {L.}~\bibnamefont {Hofstetter}}, \bibinfo {author} {\bibfnamefont {S.}~\bibnamefont {Csonka}}, \bibinfo {author} {\bibfnamefont {J.}~\bibnamefont {Nygård}},\ and\ \bibinfo {author} {\bibfnamefont {C.}~\bibnamefont {Schönenberger}},\ }\href {https://doi.org/10.1038/nature08432} {\bibfield  {journal} {\bibinfo  {journal} {Nature}\ }\textbf {\bibinfo {volume} {461}},\ \bibinfo {pages} {960} (\bibinfo {year} {2009})}\BibitemShut {NoStop}%
\bibitem [{\citenamefont {Recher}\ \emph {et~al.}(2001)\citenamefont {Recher}, \citenamefont {Sukhorukov},\ and\ \citenamefont {Loss}}]{recher_andreev_2001}%
  \BibitemOpen
  \bibfield  {author} {\bibinfo {author} {\bibfnamefont {P.}~\bibnamefont {Recher}}, \bibinfo {author} {\bibfnamefont {E.~V.}\ \bibnamefont {Sukhorukov}},\ and\ \bibinfo {author} {\bibfnamefont {D.}~\bibnamefont {Loss}},\ }\href {https://doi.org/10.1103/PhysRevB.63.165314} {\bibfield  {journal} {\bibinfo  {journal} {Physical Review B}\ }\textbf {\bibinfo {volume} {63}},\ \bibinfo {pages} {165314} (\bibinfo {year} {2001})}\BibitemShut {NoStop}%
\bibitem [{\citenamefont {Tjernshaugen}\ \emph {et~al.}(2024)\citenamefont {Tjernshaugen}, \citenamefont {Amundsen},\ and\ \citenamefont {Linder}}]{tjernshaugen_crossed_2024}%
  \BibitemOpen
  \bibfield  {author} {\bibinfo {author} {\bibfnamefont {J.~B.}\ \bibnamefont {Tjernshaugen}}, \bibinfo {author} {\bibfnamefont {M.}~\bibnamefont {Amundsen}},\ and\ \bibinfo {author} {\bibfnamefont {J.}~\bibnamefont {Linder}},\ }\href {https://doi.org/10.1103/PhysRevB.110.224502} {\bibfield  {journal} {\bibinfo  {journal} {Physical Review B}\ }\textbf {\bibinfo {volume} {110}},\ \bibinfo {pages} {224502} (\bibinfo {year} {2024})}\BibitemShut {NoStop}%
\bibitem [{\citenamefont {Wei}\ and\ \citenamefont {Chandrasekhar}(2010)}]{wei_positive_2010}%
  \BibitemOpen
  \bibfield  {author} {\bibinfo {author} {\bibfnamefont {J.}~\bibnamefont {Wei}}\ and\ \bibinfo {author} {\bibfnamefont {V.}~\bibnamefont {Chandrasekhar}},\ }\href {https://doi.org/10.1038/nphys1669} {\bibfield  {journal} {\bibinfo  {journal} {Nature Physics}\ }\textbf {\bibinfo {volume} {6}},\ \bibinfo {pages} {494} (\bibinfo {year} {2010})}\BibitemShut {NoStop}%
\bibitem [{\citenamefont {Campos}(2000)}]{campos_three-photon_2000}%
  \BibitemOpen
  \bibfield  {author} {\bibinfo {author} {\bibfnamefont {R.~A.}\ \bibnamefont {Campos}},\ }\href {https://doi.org/10.1103/PhysRevA.62.013809} {\bibfield  {journal} {\bibinfo  {journal} {Physical Review A}\ }\textbf {\bibinfo {volume} {62}},\ \bibinfo {pages} {013809} (\bibinfo {year} {2000})}\BibitemShut {NoStop}%
\bibitem [{\citenamefont {Schaeff}\ \emph {et~al.}(2015)\citenamefont {Schaeff}, \citenamefont {Polster}, \citenamefont {Huber}, \citenamefont {Ramelow},\ and\ \citenamefont {Zeilinger}}]{schaeff_experimental_2015}%
  \BibitemOpen
  \bibfield  {author} {\bibinfo {author} {\bibfnamefont {C.}~\bibnamefont {Schaeff}}, \bibinfo {author} {\bibfnamefont {R.}~\bibnamefont {Polster}}, \bibinfo {author} {\bibfnamefont {M.}~\bibnamefont {Huber}}, \bibinfo {author} {\bibfnamefont {S.}~\bibnamefont {Ramelow}},\ and\ \bibinfo {author} {\bibfnamefont {A.}~\bibnamefont {Zeilinger}},\ }\href {https://doi.org/10.1364/OPTICA.2.000523} {\bibfield  {journal} {\bibinfo  {journal} {Optica}\ }\textbf {\bibinfo {volume} {2}},\ \bibinfo {pages} {523} (\bibinfo {year} {2015})}\BibitemShut {NoStop}%
\end{thebibliography}%

\end{document}